\documentclass[preprint2]{aastex}
\usepackage{graphicx}
\shorttitle{GPU software correlation}
\shortauthors{R. B. Wayth et al.}

\begin{document}

\title{A GPU based real-time software correlation system for the Murchison Widefield Array prototype.}
\author{Randall B. Wayth\altaffilmark{1}, Lincoln J. Greenhill}
\affil{Harvard-Smithsonian Center for Astrophysics}
\affil{60 Garden St Cambridge, MA 02138 USA}
\email{rwayth@cfa.harvard.edu}
\altaffiltext{1}{Current address: Curtin Institute of Radio Astronomy. GPO Box U1987, Perth WA 6845. Australia.}
\and
\author{Frank H. Briggs}
\affil{Research School of Astronomy and Astrophysics, Australian National University}
\affil{Cotter Road Weston ACT 2611, Australia}

\begin{abstract}
Modern graphics processing units (GPUs) are inexpensive commodity hardware that offer Tflop/s
theoretical computing capacity. GPUs are well suited to many compute-intensive tasks including
digital signal processing.

We describe the implementation and performance of a GPU-based digital correlator for radio astronomy.
The correlator is implemented using the NVIDIA CUDA development environment. We evaluate three design options
on two generations of NVIDIA hardware. The different designs utilize the internal registers, shared memory and
multiprocessors in different ways. We find that optimal performance is achieved with the design that
minimizes global memory reads on recent generations of hardware.

The GPU-based correlator outperforms a single-threaded CPU equivalent by a factor of 60 for a 32 antenna array, and
runs on commodity PC hardware. The extra compute capability provided by the GPU maximises the correlation capability
of a PC while retaining the fast development time associated with using standard hardware, networking and
programming languages. In this way, a GPU-based correlation system represents a middle ground in design space between
high performance, custom built hardware and pure CPU-based software correlation.

The correlator was deployed at the Murchison Widefield Array 32 antenna prototype system
where it ran in real-time for extended periods. We briefly describe the data capture, streaming and correlation
system for the prototype array.

\end{abstract}

\keywords{
Astronomical Instrumentation, Astronomical Techniques
}

\section{Introduction}
The use of interferometric techniques in radio astronomy allow astronomers to achieve extraordinary angular resolution
without having to build large monolithic antennas.
Interferometric radio telescope arrays create a virtual antenna of large diameter by placing many relatively small
antennas over a large area (anywhere from 100s of meters to 1000s of kilometers).

Interferometric radio telescopes work by correlating the signals between antennas in the array to form complex-valued
correlation products, or `visibilities', which are accumulated over a short time then
stored for subsequent calibration and image processing.
The visibilities are discrete samples of the Fourier transform of the sky intensity as seen by each
antenna in the array.
An image of the sky can be made by placing the calibrated visibilities onto a grid, Fourier transforming the grid
then deconvolving the resulting image.
More details of the fundamentals of radio interferometry can be found in one of the many texts
on the subject \citep[e.g.][]{2001isra.book.....T}.


The Murchison Widefield Array (MWA) is a new low-frequency radio telescope that is currently under construction
at the Murchison Radio Observatory in Western Australia. The full array will consist of 512 antennas
spread over approximately 1km. The array has some ambitious science goals including measurement of the power
spectrum from the Epoch of Reionization (EoR) in the early universe, measurement of properties of the
heliosphere and discovery and monitoring of transient and variable radio sources
\citep{2005SPIE.5901..124S,2006ApJ...638...20B}.

During 2008, a 32 antenna prototype system was deployed in Western Australia.
The full MWA will have a dedicated hardware correlator, however to enable interferometry for the prototype system
in real-time, a GPU-based software correlator was developed.

In this paper, we describe the GPU-based software correlation system that was deployed with the 32 antenna
prototype of the MWA.
In \S \ref{sec:corr} we provide a brief overview of the requirements for radio telescope array correlators. 
We describe the prototype system and the data capture hardware in \S \ref{sec:mwa32t}.
In \S \ref{sec:gpu_corr}, we review the basic design requirements and issues for an FX correlator system on a GPU.
In \S \ref{sec:perf}, we list the performance of the GPU correlator for the prototype system
and compare to a conventional CPU-based correlator.
In \S \ref{sec:discussion} we discuss the performance of the correlator designs in terms of the theoretical
peak performance and their utilisation of GPU resources.
We also discuss some potential applications for similar signal processing tasks in radio astronomy.

\section{Correlators}
\label{sec:corr}

Radio telescope correlators form visibilities from Nyquist-sampled band-limited signals received from each antenna.
Signals from all antenna pair combinations are correlated and integrated over some time,
each forming a complex-valued correlation coefficient, or visibility.

For a correlator with $N_I$ input signals there are $N_I(N_I-1)/2$ cross-correlation products (between different
input signals) and $N_I$ auto-correlation products, which is the correlation of a signal with itself. The total number
of correlation products for $N_I$ inputs, including auto-correlations, is $N_I(N_I+1)/2$.

All modern correlators used in radio astronomy are digital.
Once a band-limited signal is digitized, correlation becomes a signal processing problem and
there is some flexibility in the choice of hardware and software that performs the task.
Historically, large radio telescope facilities have custom-built hardware correlators based on ASICs or FPGAs.
More recently, software based correlation systems built on clusters of commodity computers have been used
for modest bandwidth applications, e.g. at the GMRT \citep{2008_GMRT_Roy}, or for processing of VLBI data
\citep{2007PASP..119..318D}. Such systems offer the benefits of using commodity computing hardware, networking
and high-level programming languages.

Astronomers also often implement a spectrometer within the correlator to
study the details of the spectrum contained within the sampled band.
In the MWA application, fine frequency resolution is required for the implementation of
delay and phase calibration of the array, as discussed below.
The correlator is therefore required to
divide the input bandwidth into consecutive sub-bands, or frequency `channels', and form the correlation
between inputs for each channel.

Software correlation systems are well suited to the so-called `FX' correlator design.
Conceptually, an FX correlator takes batches of samples from two antennas,
Fourier transforms (`F') the signals into a frequency spectrum
then then does a channel-by-channel multiplication (`X') of the first spectrum with the complex conjugate of the second
to form the correlation product, the complex-valued cross-power spectrum.
The results of the cross-multiplications are accumulated for each baseline.
The samples input to the correlator may be real or complex depending on the data capture system and any pre-processing.
For real-valued samples, $2N_C$ samples are transformed into $N_C + 1$ spectral channels, including the DC term.
For complex valued samples, $N_C$ samples are transformed into $N_C$ spectral channels.
As noted later, the MWA receiving system produces complex-valued samples, which will be assumed henceforth.

The cross-multiply is performed for all antenna-pair combinations, hence the
number of operations in an FX correlator is dominated by the `X' stage when the number of antennas is large.
Specifically, the number of operations to perform the FFT is $O \{ \log_2(N_C) \}$ \emph{per sample},
which is effectively constant,
whereas the number of operations to perform the cross-multiply is $O \{ N_I \}$ per sample.
Since the number of samples to process per second scales as $ N_I \times B$ where $B$ is the bandwidth of the data,
an FX correlator must perform $O \{B N_I (N_I + \mathrm{const}) \}$ operations per second.

Many correlators are also required to
apply a delay to the inputs from the antennas to compensate for the geometric delay between antennas.
The delay depends on the direction on the sky that the telescope is pointing, which can add
significant additional complexity at the input to the correlator.
For the MWA, the delay correction is applied post correlation as a phase correction without loss of sensitivity.
This can be done because the MWA's baselines are short enough and the correlated spectral resolution is fine enough
that the phase gradient within a channel due to geometric delay is negligible.
This simplifies the requirements of the correlator.

\section{The MWA 32 antenna prototype}
\label{sec:mwa32t}

The MWA prototype consists of 32 antennas placed in an approximate Reuleaux triangle configuration
with maximum separation between antennas approximately 350 meters. Each antenna comprises 16 dual linear polarization
dipoles operating as a phased array.
The signals from the dipoles are combined in an analog beamformer, that electronically steers the antennas
to the desired pointing direction by applying the appropriate delay to each dipole.
Each produces one analog signal for each polarization.

Signals from groups of eight antennas are fed into a receiver node. The receiver performs several
tasks, including band-limiting and adjusting the power levels of the signals \citep{2008_MWA_overview}.
The receivers internally use a polyphase filterbank to create 256 output channels, covering the 
frequency spectrum between 1.28 MHz and 327.68 MHz, each having 1.28 MHz bandwidth.
Complex valued samples from each of these channels are produced at 1.28 Msamples/second, which is the Nyquist
rate for a 1.28 MHz band.
In the prototype system, one channel is input to the correlator.

For the prototype, there are four `receiver' units, each servicing eight antennas.
Each receiver feeds data to a PC-based data capture and recording system \citep{2002ivsg.conf..128R}.
The receivers are locked to a common 655.36 MHz clock so that samples recorded on each data capture PC are
from the same instant, preserving interferometric phase across the array.
The recorded samples are represented by 5 bits each for the real and imaginary part of the number
in two's-complement format.
For convenience in the prototyping effort, these 10-bit samples were stored as 16-bit words.
The format of the recorded data is in groups of 17 2-byte words, consisting of a marker/counter word followed by 16
data words, one for each input to the receiver (8 antennas $\times$ 2 polarizations).
The net data rate per capture PC is 43.52 MB/s.

Although the data recording systems are running on independent PCs, the captured samples are explicitly synchronized
by the common clock in the receivers. A relatively simple data aggregation system is used to
stream the data to a common PC with a GPU for correlation. This aggregation system is not synchronized; it
merely needs to preserve the number and ordering of samples, which is achieved using standard TCP/IP networking.
The aggregation system removes the marker words and interleaves the four streams so that the 64 samples for
the same time instant are grouped together for input to the correlator.

\section{GPU correlator design}
\label{sec:gpu_corr}

\begin{figure*}
\includegraphics[scale=0.6]{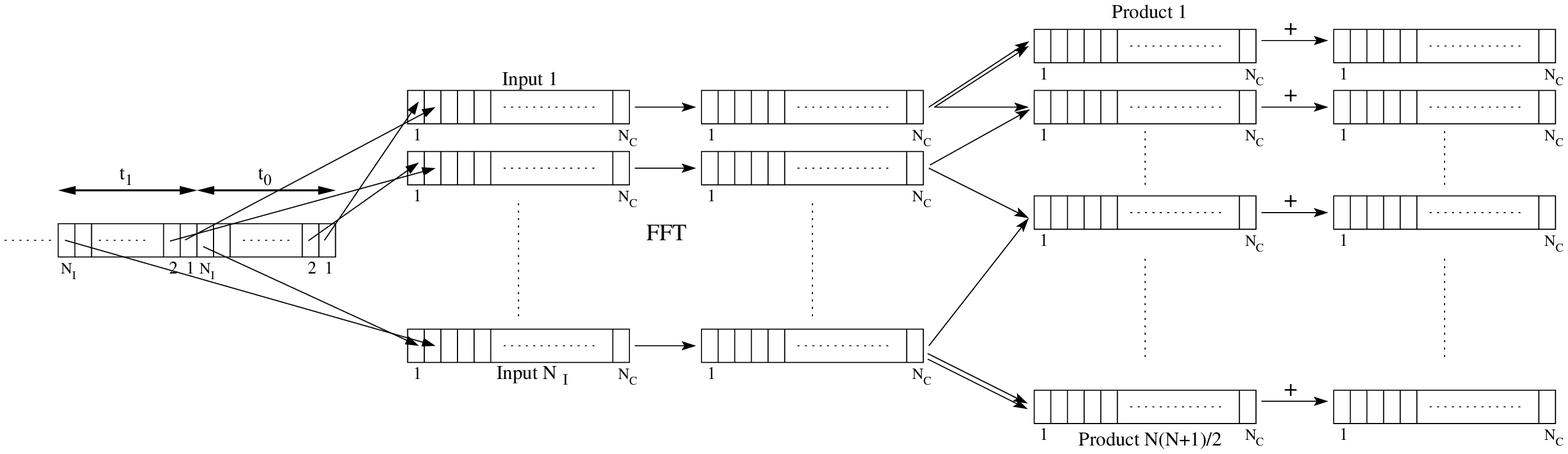}
\caption{Basic design for an FX correlator with complex input samples.
A batch of $N_C\times N_I$ samples, ordered by time, is reordered into $N_I$ blocks of $N_C$ samples for FFT.
Correlation products are formed by multiplying one spectrum with the conjugate of another, including auto-correlations.
The resulting correlations are added into accumulators.
After a specified time, usually a few seconds depending on the requirements of the astronomer,
the accumulators are read and divided by the number of times they were updated, to form a time-averaged correlation product.
They are then reset for the next time interval.}
\label{fig:fx}
\end{figure*}

We implemented the correlator using NVIDIA GPU hardware and CUDA\footnote{http://nvidia.com/cuda} programming environment. 
There are three basic steps in the correlation process:
\\1- unpacking a block of data, which involves reordering samples as discussed below,
\\2- Fourier transforming into frequency space, and
\\3- performing the cross-multiplication of spectra and adding the results into accumulators.
The correlator is configured to
receive input streams from $N_{I}$ inputs
(where here $N_I = 64 = 2 $ polarizations $\times$ 32 antennas), and form
correlation products with $N_{C}$ spectral channels. The basic data flow is shown in Fig. \ref{fig:fx}.

Code in CUDA is written in C, with some language-specific extensions.
Optimal use of a GPU requires some understanding of how the GPU operates at a relatively low level and for the
software design to exploit multi-threaded SIMD-like hardware. We briefly review the most important concepts here.

Code on an NVIDIA GPU is executed by one of many parallel multiprocessors, each of which contains eight compute cores.
A multiprocessor runs code in blocks of threads, each thread ideally executes the same instructions on different
data. GPUs have global memory and registers, similar to conventional computers. Registers are fast to access
but are only available within a thread. Global memory can be accessed by all threads but has high latency
and there is no cache for general global memory.
Some data may also be shared between the threads through the use of \emph{shared memory}. GPU shared memory does not
have an analog on conventional CPUs. It is like a register that can be accessed by a small group of threads.
Accessing shared memory is much faster than global memory.
The GPU also has a small amount of read-only \emph{constant memory} which is cached and fast to access.

There are four main design considerations for efficient use of the GPU:
\\1- Coalesced global memory access. If consecutive threads in a thread block access consecutive locations in 
global memory, the memory operation will be coalesced into a single operation by the GPU memory management hardware.
Using non-coalesced operations can have a significant performance penalty.
\\2- Using shared memory to reduce global memory operations. Data that can be shared between threads need only be read
once from global memory then is accessible to a group of threads through shared memory.
\\3- Register and shared memory use. The total number of registers and shared memory required by a thread dictates
how many blocks of threads can be run concurrently on the GPU.
The fraction of the GPU compute resources actually used is the `occupancy'.
\\4- Thread serialization. Designs that use shared memory to reduce global memory access
may require forced synchronisation points in the code
to guarantee all required data have been loaded before continuing. This synchronisation
can force groups of threads to wait, which reduces the efficiency of parallel processing.

Unpacking the data requires taking a block of $N_{I} \times N_{C}$ complex samples which are ordered with
input varying most quickly, converting to floating-point format and putting the samples into $N_{I}$ arrays of
length $N_{C}$, with channel number varying most quickly (Fig \ref{fig:fx}, left two columns).
With the exception of converting to floating-point, this is equivalent to a matrix transpose
for the block of samples. This will force non-coalesced memory reads,
although as we note later, the unpacking operation is a small fraction of the total
processing time for the MWA prototype application even when using non-coalesced operations.

The second step is performed by the built-in FFT library in CUDA. For the complex valued
samples of the MWA, the FFT is $N_{C}$ channels in input and output (second and third columns of Fig. \ref{fig:fx}).
This results in coalesced memory operations for power-of-two transforms.
The correlator also supports real-valued input samples.
For real valued samples, $2N_{C}$ samples are transformed into $N_{C}+1$ channels.
To avoid having odd-sized array lengths for real-valued
data and the associated memory coalescence problems, we also use complex-to-complex transform of size $2N_{C}$
for real-valued samples and ignore the redundant half of the result.

\begin{figure}
\plotone{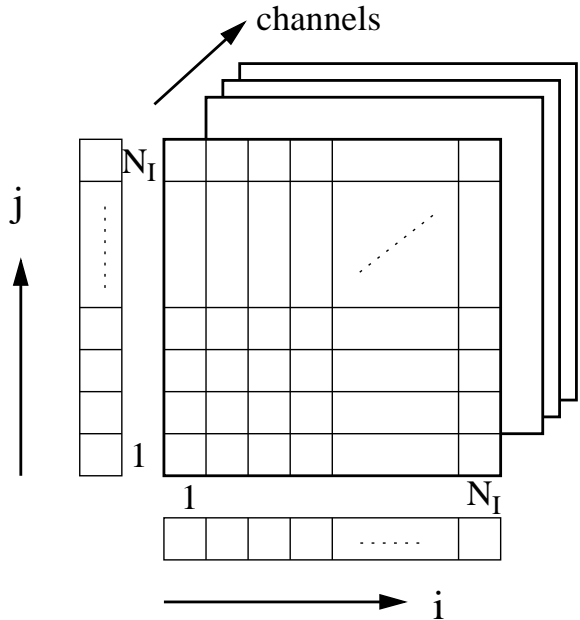}
\caption{Correlation products can be represented by the outer product of a vector with its complex conjugate
to form a matrix. Elements in the vector are samples from each input in frequency space for a frequency channel.
Auto-correlations are formed along the diagonal with redundant copies of the cross-correlation
products in the top and bottom halves. The products are formed independently for each frequency channel.}
\label{fig:corr_matrix}
\end{figure}

The third step performs the cross-multiplication and accumulation (CMAC).
The correlation products for a given frequency channel can be represented in a matrix of size $N_I \times N_I$
formed by the outer product of a vector with its complex conjugate as shown in Fig \ref{fig:corr_matrix}.
The vector contains the samples (in frequency space) for all inputs in a channel for a single time instant.

\citet{2008ExA....22..129H} (hereafter HHS08) examined various design options for
optimizing GPU performance for the correlation task. Here, we evaluate three design
options on two generations of hardware. Two of our designs are very similar to
the `pair parallel' and `group parallel' approaches in HHS08, and we reuse the names.
A third design, not considered in HHS08,
we call the `hybrid group' because its design has elements of the first two.

The \emph{pair parallel} (PP) approach is conceptually the simplest. $N_I(N_I+1)/2$ blocks of threads are created, with
$N_{C}$ threads in each block. Each thread is independent of all others and computes a single correlation product
for a single channel and adds it into an accumulator. Hence a block of threads computes all channels for one
correlation product.
For a large number of frequency channels, it is straightforward to divide
$N_{C}$ into chunks that optimise the GPU occupancy (e.g. 128 channels)
by creating $N_I(N_I+1)/2 \times N_{C}/128$ thread blocks.
This approach does not use shared memory. Each correlation product requires two reads from global memory per channel
and requires one accumulator per thread. Our pair parallel design differs from HHS08 only in the number of thread
blocks created. Instead of creating $N_{I}^2$ threads per channel
and using a condition to test for the redundant combinations, our design creates $N_{I}(N_{I}+1)/2$ thread blocks.
This design requires the threads to decode which input pair they correspond to.
Instead of using expensive modulo operations to decode the input pair,
we pre-compute a lookup table for this task and load it into the GPU constant memory, which is cached and fast to
access in our application because all threads in a thread block are reading the same data from the constant memory.

The \emph{group parallel} (GP) approach aims to minimise the reads from global memory by sharing common data
between a small group of threads, where $G$ is the size of the group.
Each block of threads runs $128/G$  (where $G$ = 2 or 4) channels in the frequency direction and $G$ inputs in the $i$
direction. A total of $(N_I/G)(N_I/G + 1)/2 \times N_CG/128$ thread blocks are created,
with redundant input pairs doing no work. Each thread loads one value
from the $i$ axis, which is shared, then loads one value from the $j$ axis which is not shared. The thread then
calculates the $G$ products along the $i$ axis and adds them to accumulators. This approach uses one complex number
(8 bytes) of shared memory per thread and $G$ accumulators per thread.
The design makes $2G$ reads from global memory to calculate $G^2$ products.
Because data are shared between threads, synchronisation is required.

The \emph{hybrid group} (HG) approach aims to use shared memory to reduce the total number of global memory reads
while maximising occupancy and avoiding thread synchronisation. In this approach, $N_I \times N_I/G$ (G=2 or 4) thread
blocks are created, including a redundant half of the correlation matrix.
Thread blocks for the redundant half do no work and exit.
Each thread loads the value from the $i$ axis into shared memory, then loads $G$ successive values
from the $j$ axis to calculate $G$ correlation products. $G+1$ reads are required to calculate $G$ products.
Because data are not shared between threads, there is no need for thread synchronisation. This design simply uses shared
memory as additional register space. 

In all designs, consecutive threads process consecutive channels so all memory operations are coalesced.
Additionally, all designs transfer and process large batches of data per call to the GPU.
This has the benefit that the correlation products can be accumulated in registers, thereby reducing the total
number of fetch-add-store global memory accesses, and by reducing the overhead of making the GPU calls.

For the designs that use shared memory, the choice of $G=2$ or $4$ is pragmatic. $G$ must divide evenly into $N_I$
hence $G$ must be even for our system. For $G > 4$, the designs require an increased number of registers
that significantly reduces the GPU occupancy.

\section{Performance}
\label{sec:perf}

The performance of the designs was tested on three different NVIDIA GPUs, representing two generations of hardware
with different capabilities, as shown in Table \ref{tab:gpu_list}.

\begin{table}
\begin{centering}
\caption{\small GPU models used for performance testing.}
\begin{tabular}{lllll}
GPU model         & MP\tablenotemark{a}    & Reg\tablenotemark{b}     & BW\tablenotemark{c}    \\ \hline
GeForce 8800GTS   & 12    & 8192    &  64   \\
GeForce 8800GTX   & 16    & 8192    &  86.4 \\
Tesla C1060       & 30    & 16384   &  102
\end{tabular}
\tablenotetext{a}{MP = number of multiprocessors on that model.}
\tablenotetext{b}{Reg = the number of 32-bit registers available per thread block.}
\tablenotetext{c}{BW = the nominal bandwidth to GPU global memory (GB/s).}
\label{tab:gpu_list}
\end{centering}
\end{table}

\begin{table*}
\caption{Execution configuration for different CMAC designs and occupancy for the different model GPUs.}
\begin{center}
\begin{tabular}{lllcccc}
           &                             &               &        &         & \multicolumn{2}{c}{Occupancy} \\
Design     & Grid size                   & TB size       & N Regs &  ShMem  &  8800 &  C1060 \\ \hline
PP         & $N_I(N_I+1)/2$              & 128           & 12     &  88     & 5/6       & 1 \\
GP $(G=2)$ & $2 \times N_I/2(N_I/2+1)/2$ & $64 \times 2$ & 14     &  1120   & 2/3       & 1 \\
GP $(G=4)$ & $4 \times N_I/4(N_I/4+1)/2$ & $32 \times 4$ & 16     &  1120   & 2/3       & 1 \\
HG $(G=2)$ & $N_I \times N_I/2$          & 128           & 15     &  1120   & 2/3       & 1 \\
HG $(G=4)$ & $N_I \times N_I/4$          & 128           & 20     &  1120   & 1/2       & 3/4 
\end{tabular}
\end{center}
\tablecomments{
All designs generate 128 frequency channels. PP = pair parallel, GP = group parallel, HG = hybrid group.
Columns 2 and 3 show the number of thread blocks (TB) created and the dimensions of each thread block.
Columns 4 and 5 show the number of registers used per thread and the total amount of shared memory (bytes) used in
a thread block.
Columns 6 and 7 show the occupancy of each multiprocessor on the different generations of hardware.
The 8800GTS and 8800GTX are the same generation hardware, so have the same occupancy.}
\label{tab:design_specs}
\end{table*}

As noted in \S \ref{sec:gpu_corr}, the performance of the correlator will be mostly determined by the use of global
memory, occupancy and thread synchronisation.
Table \ref{tab:design_specs} shows the execution configuration of the GPU CMAC kernels and the
multiprocessor occupancy for the different designs on the different hardware. The occupancy is a
discontinuous function of number of registers used, the amount of shared memory used and the number of
threads in a thread block. The additional registers available in the C1060 means that multiprocessors are
fully occupied for all but the HG $(G=4)$ design.

Real-time data processing for the MWA 32 antenna system requires processing data from 4 receivers with 1.28MHz
bandwidth. Each receiver streams data at a rate of 40.96 MB/s (1.28MHz * 8 antennas * 2 polarizations * 2 bytes/sample), 
so the total data rate into the correlator is 163.84 MB/s.
We also implemented a ``1-byte'' mode where the data samples
were reduced from 2-byte words to a single byte, at the cost of 1 bit of precision in the real and imaginary parts
of the sample. The 1-byte mode halves the correlator input volume, and reduces the total
amount of global memory reads in the unpack phase, but otherwise does not
change the amount of work the GPU must do for the FFT and CMAC.

The system was configured to correlate 128 spectral channels, therefore
16384 bytes (64 inputs *128 channels *2 bytes/sample) are required per unpack-FFT-CMAC cycle.
The data were transferred to the GPU and processed in batches of $32 * 16384$ bytes to reduce the effect of overhead
incurred when transferring data or executing tasks on the GPU. We found that for batch sizes greater than around 20,
the total time taken to run the correlator did not improve, indicating that the transfer/execution overhead for
such large batches was negligible.

\begin{table*}
\begin{center}
\caption{\small Breakdown of the fraction of time spent on the main tasks in the correlator to process one second of data
for the MWA 32 antenna system (64 inputs, 1.28MHz complex samples, 128 frequency channels).}

\begin{tabular}{lrrr}

Task                                      &  \multicolumn{3}{c}{time (ms)}         \\ \hline \hline
Unpack data - CPU\tablenotemark{a}        &  \multicolumn{3}{c}{1267}         \\ 
FFT  - CPU                                &  \multicolumn{3}{c}{1840}         \\ 
CMAC - CPU                                &  \multicolumn{3}{c}{20950}         \\ \hline
Stage data in host memory\tablenotemark{b}&  \multicolumn{3}{c}{117}     \\
Transfer data to GPU                      &  \multicolumn{3}{c}{57}      \\           \hline
                                          & 8800GTS & 8800GTX & C1060  \\ \hline
Unpack data                               &  70     &    51   &  53  \\
FFT                                       &  141    &    92   &  53  \\
CMAC - PP                                 &  906    &   640   &  494 \\
CMAC - GP $G=2$                           &  635    &   432   &  268 \\
CMAC - GP $G=4$                           &  519    &   343   &  185 \\
CMAC - HG $G=2$                           &  785    &   565   &  393 \\
CMAC - HG $G=4$                           &  675    &   487   &  340 
\end{tabular}
\tablenotetext{a}{Unpacking on the CPU version includes reading the data.}
\tablenotetext{b}{Data staging on the host PC is performed while the CMAC operation for the previous batch of data is running
on the GPU, so does not contribute to total time.}
\label{tab:perf}
\end{center}
\end{table*}

For testing and performance comparison we also implemented a single-threaded CPU version of the correlator.
The CPU version of the code is written in C using single precision floating point types and uses the FFTW v3 library.
It was compiled with standard compiler optimization settings.
The correlators were run on a workstation running 64-bit CentOS 5 with CUDA v2.1.
The workstation contains a dual-core AMD Opteron 2.2GHz processor, 2GB RAM, and a Tyan S2925 motherboard.

Table \ref{tab:perf} shows the amount of time spent in each section of the code when
processing one second of data for the MWA prototype using 2-byte samples. As expected, the majority
of the time is in the CMAC operation. The CPU-based code runs over 20 times slower than real-time
whereas the GPUs complete all tasks in real-time or better.
By transferring data to the GPU in large batches,
the transfer rate approached 3 GB/s including overhead, which is close to the peak theoretical performance
of 4 GB/s for the 16-lane PCI express v1.0 bus on the motherboard.

\begin{figure*}
\centering
\includegraphics[width=9cm,height=15cm,angle=-90]{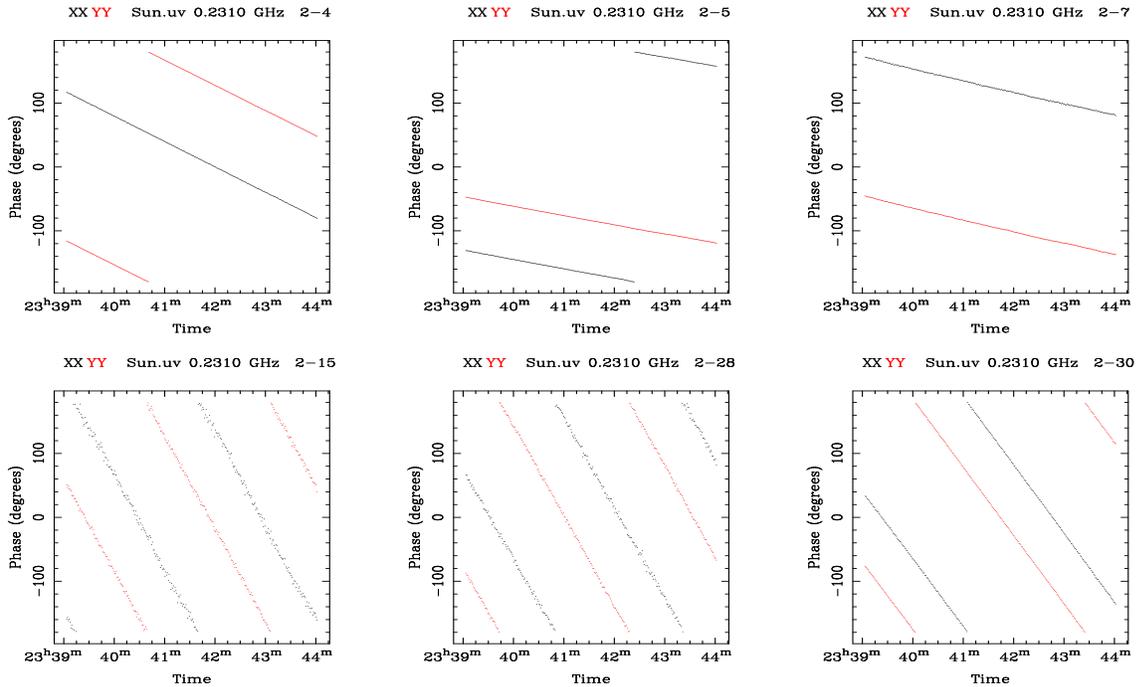}
\caption{Example raw (uncalibrated) interferometric fringes from observing the Sun for 5 minutes at 230 MHz.
Only some baselines are shown. Clear interferometric fringes are visible as time varying phase.
The top row contains relatively short baselines ranging from 50 to 100 meters, the bottom row contains relatively
long baselines between 250 and 300 meters. The rate of change of visibility phase depends on the baseline
length and orientation.}
\label{fig:example_fringes}
\end{figure*}

The GPU correlation system was deployed on site during an engineering test run in November 2008. 
The system was configured to stream 1-byte data from the four data capture PCs to a fifth, equipped with GPU, via
standard gigabit ethernet.

The telescope was configured to track astronomical sources for 5 minute intervals, then to change frequency and/or
source. This required the correlator to run in real-time for 5 minutes continuously with short breaks between.
The real-time system performed as expected, with successful correlation on several known strong radio sources
at several frequencies including the Sun.
Interferometric fringes from the Sun in two polarisations (XX and YY)
at 230MHz for some example baselines are shown in Fig. \ref{fig:example_fringes}.

\section{Discussion}
\label{sec:discussion}

Commodity hardware allows data to be streamed at $\sim 0.8$ Gb/s using standard gigabit ethernet. It is common
in radio astronomy for samples to be represented by as few as 2 bits, hence commodity PCs can receive
continuous streams with 
hundreds of Msamples/sec. Recall that to FFT the data, $O \{ \log_2(N_C) \}$ operations are required per sample
and $O \{ N_I \}$ are required to cross-multiply, where an operation is a complex multiply and add in both cases.
If data are streamed at hundreds of Msamples/sec,
a single thread on a GHz-class CPU can perform only $\sim 10$ operations per sample unless SIMD instructions are used.
For arrays with many antennas, the CMAC task requires too many operations per sample for a single CPU thread.
The work must be accomplished via a multi-threaded program in an SMP machine or by offloading the work to
another device such as a GPU. For the CMAC task, GPUs are well suited due to the high arithmetic intensity per sample.

An additional benefit of performing the software correlation task on the GPU is that the correlator itself creates only
a small load on the host PC. This is important because the data aggregation task accepts streamed data from two
gigabit ethernet ports at $81.92$ Msamples/sec (total). The relatively simple task of aggregation still requires several 
operations (reads and copies) per sample, resulting in significant load on the host system just to prepare the data
for correlation.

To form the correlation products in real-time, the GPU must execute $B N_I(N_I + 1)/2$ CMACs per second,
where $B$ is the bandwidth of data.
This involves fetching $\epsilon \times 4 B N_I(N_I + 1)$ bytes/sec from global memory (single precision complex
floating point data)
where $\epsilon$ is the aggregate number of global memory reads per channel per product and is $2,1.5,1.25,1,0.5$
for the PP, HG $G=2$, HG $G=4$, GP $G=2$ and GP $G=4$ designs respectively.
If the computation is memory bandwidth limited and $M$ is the throughput of the GPU global memory,
the maximum theoretical signal bandwidth that the GPU can support is
$B = M/(4 \epsilon N_I(N_I + 1) )$ for the CMAC operation.

HHS08 reported best performance with the group parallel design with $G=4$ running on an 8800GTS.
We confirm this result on all three of the GPUs that were tested. 
Because all designs perform virtually the same amount of arithmetic (there will be small differences due to
array indexing calculations, loop control and redundant product calculations),
we can deduce where the bottlenecks exist by comparing the execution times among designs.


\begin{table*}
\begin{center}
\caption{Actual/theoretical signal bandwidth (MHz) that can be processed by different correlator designs}
\begin{tabular}{lccccc}
GPU model         & PP     & HG $G=2$ & HG $G=4$ & GP $G=2$ & GP $G=4$     \\ \hline
GeForce 8800GTS   & 1.4/1.9 (73\%) & 1.6/2.6 (64\%) & 1.9/3.1 (62\%) & 2.0/3.8 (52\%) &  2.5/7.7 (32\%) \\
GeForce 8800GTX   & 2.0/2.6 (77\%) & 2.3/3.4 (65\%) & 2.6/4.2 (63\%) & 3.0/6.1 (57\%) &  3.7/10  (36\%) \\
Tesla C1060       & 2.6/3.1 (85\%) & 3.3/4.1 (80\%) & 3.8/4.9 (77\%) & 4.8/6.1 (78\%) &  6.9/12 (56\%)
\end{tabular}
\tablecomments{This is for a 64 input system, assuming the operation is memory bandwidth limited. The numbers in parentheses
show the fraction of theoretical performance that was achieved.}
\label{tab:theoretical_perf}
\end{center}
\end{table*}

Table \ref{tab:theoretical_perf} shows the actual and theoretical bandwidth for the CMAC operation 
that can be processed by the different GPUs for a 64 input system,
assuming the computation is memory bandwidth limited, for the different correlator designs.
The actual performance is calculated using the execution times from Table \ref{tab:perf},
and therefore includes all the overhead of making the call to the GPU from the host system and within the GPU itself,
but does not include the cost of transferring, unpacking or FFTing the data.
The designs that access global memory the least have the greatest theoretical bandwidth.
The numbers in parentheses are the ratio of actual/theoretical signal bandwidth, expressed as a percentage,
and represent how much time each design spends in global memory access operations.
A small number indicates that the GPU is spending time \emph{not} waiting for global memory, which means that
it is doing arithmetic and hence is being efficiently utilized.

The execution time of the PP
design in Table \ref{tab:theoretical_perf} is close to the theoretical performance.
This indicates that the basic PP design is memory bandwidth limited, with 2 global memory fetches per complex
multiply and add, despite having the highest occupancy.
By reducing the total load on global memory, the HG and GP designs run significantly faster than the PP design.

The HG $G=4$ and GP $G=2$ designs have a similar load on global memory
with $5/4$ and $1$ fetches per CMAC respectively. The small fractional improvement in execution time
and lower performance relative to peak of the GP $G=2$, between the two
designs on the 8800 models, does not match the naive estimate of potential improvement from memory bandwidth alone.
This indicates that thread synchronisation is forcing the multiprocessors to wait on the 8800 models,
offsetting the reduced memory usage and increased multiprocessor occupancy of the GP design.
On the C1060, however, the bottleneck appears to be eliminated with the HG $G=4$ and GP $G=2$
designs running at almost the same fraction of theoretical peak.

The low memory throughput compared to theoretical for the GP $G=4$ design indicates that the CMAC
operation is no longer limited by memory bandwidth and is therefore spending much of the time
performing the arithmetic. We conclude that this design is using the GPU processing resources optimally.

\subsection{Further work}
As a signal processing device, GPUs compare favorably to custom-built hardware both from the processing power
and development time perspective. The development is more like traditional software engineering, with a C-like
programming language, while the performance is more like custom hardware.
There are limitations to the scalability of the correlator, however.
The amount of work the correlator must perform per time interval scales as $B N_I(N_I+1)/2$.
The $N_I^2$ scaling with inputs is the limiting factor for radio arrays with large numbers of antennas.
For a given bandwidth, all of the CMAC work could in principle be divided over several GPUs.
Such a setup would require all of the input data to be streamed to all GPUs and for
all GPUs to do all FFTs. Then the CMAC would be divided between GPUs.
Given the speed of modern PCI-express buses relative to gigabit ethernet, this scheme would probably be limited
by the number of GPUs that can be attached to a single host.

A potential improvement to the GPU correlator would be to use a polyphase
filterbank (PFB) instead of an FFT when forming the spectra from the time-series data.
FFTs have poor performance for narrowband signals that do not fall exactly in the center of a spectral channel,
the resulting signal power is spread over several adjacent channels. This spectral leakage is undesirable for
high signal-to-noise or high-precision experiments.
Using a PFB can greatly reduce spectral leakage, with the extra cost coming in the unpack stage of the correlator.

\section{Conclusions}
We have evaluated several designs for a GPU-based radio astronomy correlator system using NVIDIA's hardware and CUDA programming
environment. The designs were compared for the MWA's 32 antenna prototype system which must correlate 64 data streams
of 1.28MHz bandwidth into 128 spectral channels in real-time.

We found the group parallel design, which uses the GPU shared memory to minimise the amount of global memory reads,
was the optimal solution on three different models of GPU hardware. Running on the current generation C1060 hardware,
the GPU correlator runs approximately 68 times faster than a single-threaded CPU equivalent and leaves the host
system resources free to perform the necessary streaming of data

\acknowledgements
RBW is grateful to Kevin Dale, Chris Harris and Chris Phillips for their thoughts and suggestions
on the design of various parts of the system.

We thank the hard work and dedication of the November 2008 site visit team:
David Basden,
Roger Cappallo,
Mark Derome,
David Herne,
Colin Lonsdale,
Merv Lynch,
Daniel Mitchell,
Miguel Morales,
Ed Morgan,
Anish Roshi,
Steven Tingay,
Mark Waterson,
Alan Whitney,
Andrew Williams
and Jamie Stevens for remote support.

We thank the team at Raman Research Institute led by Anish Roshi for enabling the data capture system.
The team includes
Srivani K S,  Prabu T., Deepak Kumar, Gopala Krishna,  Kamini P. A. and  Madhavi S.

This work was in part supported by grants AST-0457585 and PHY-0835713 of the National Science Foundation. 

Facilities: \facility{Murchison Widefield Array}.


\label{lastpage}
\end{document}